\documentclass[oldversion]{aa} 
\usepackage{natbib}
\usepackage{graphicx}
\usepackage{amsmath}
\usepackage{txfonts}
\begin{document}
   \title{{\it Chandra} observation of Cepheus A:\\ The diffuse emission of HH~168 resolved}
   \titlerunning{The diffuse X-ray emission of HH~168 resolved}

   \author{P. C. Schneider
          \and
          H. M. G\"unther
          \and
          J. H. M. M. Schmitt
          }

   \institute{Hamburger Sternwarte,
              Gojenbergsweg 112, 21029 Hamburg\\
              \email{cschneider/mguenther/jschmitt@hs.uni-hamburg.de}
             }

   \date{Received 16.07.2009 / accepted 25.08.2009}

  \abstract
   {X-ray emission from massive stellar outflows has been detected in several cases. We present a \textit{Chandra} observation of HH~168 and show that the soft X-ray emission from a plasma of 0.55~keV within HH~168 is diffuse. The X-ray emission  is observed on two different scales: Three individual, yet extended, regions are embedded within a complex of low X-ray surface brightness. Compared to the bow shock the emission is displaced against the outflow direction. We show that there is no significant contribution from young stellar objects (YSOs) and discuss several shock scenarios that can produce the observed signatures. We establish that the X-ray emission of HH~168 is excited by internal shocks in contrast to simple models, which expect the bow shock to be the most X-ray luminous.}
   \keywords{stars: winds, outflows - X-ray: ISM -  Herbig-Haro objects - ISM: jet and outflows - ISM: individual objects: HH 168}

   \maketitle
%


\section{Introduction}
Star formation is intimately linked to accretion and outflow phenomena. Early in the formation process, the collapsing proto-star is deeply embedded and therefore invisible, yet, powerful outflows emerge from these systems. At a later stage, the accretion proceeds from a disk. For low-mass stars, the magnetically-funneled infall model explains many observed phenomena successfully, but, even in their case, the jet and wind driving mechanism remains elusive. Current theories expect mass loss to be driven by magneto-centrifugal disk winds, possibly with a stellar contribution. The physical mechanism, which accelerates and collimates the outflow also remains elusive. Jets have been observed at all stages of star formation over a wide mass range in different wavelength regions. \object{HH~2} \citep{Pravdo_2001N} and \object{HH~154} \citep{2002A&A...386..204F,2003ApJ...584..843B,2006A&A...450L..17F} were the first massive, large-scale outflows discovered in X-rays. The latter object has already been observed twice, so that the proper motion of its X-ray emitting regions could be tracked. The  proper motion of 500~km~s$^{-1}$ can explain the observed temperatures in terms of simple shock models of a thin gas ramming into the ambient medium.
Examples of further evolved objects are the low-mass star \object{DG Tau} \citep{Guedel_2008, Schneider_2008, Guenther_2009}, where the X-ray emission from the jet is confined to the inner region of a few \mbox{hundred AU} and the intermediate mass Herbig~Ae/Be star \object{HD 163296} \citep{Swartz_2005,HD163296}. Because of the low surface brightness of these jets and a lack of nearby young intermediate and high-mass stars, X-rays from their outflows have only been detected in very few cases. 

Most models of high-energy emission in outflowing material adopt some type of shock.
The interface between the outflowing material and the ambient medium seems to be a plausible heating source but evidence of additional X-rays produced close to the star is increasing. Thus, a more complex scenario might be required to explain the observed phenomena. Internal working surfaces within the outflow might be responsible for the heating, but they require rapidly moving components within the flow so that the velocity difference still provides enough energy to produce X-rays.

In the case of DG~Tau, such a high velocity component has not been detected in UV, optical or IR observations. However, when compared to the outflow's total mass-loss this speculative high velocity component would constitute only a very small fraction of the total mass-loss and energy and, therefore, might have escaped detection \citep{Schneider_2008, Guenther_2009}.

This paper deals with the Cep~A West region, where the outflow \object{HH 168} is located. The central region of the complex, which presumably contains the jet driving source, will be analysed in a forthcoming article \citep{Schneider_Cep_A_main}.
In the next section, we will provide a brief overview of the Cep~A region and in the subsequent two sections the observation and results are presented. We then discuss the properties of the emitting plasma before we investigate possible scenarios explaining the observed X-ray emission.

\section{The Cepheus~A high-mass star-forming region}
\begin{figure}
  \centering
   \includegraphics[width=0.49\textwidth]{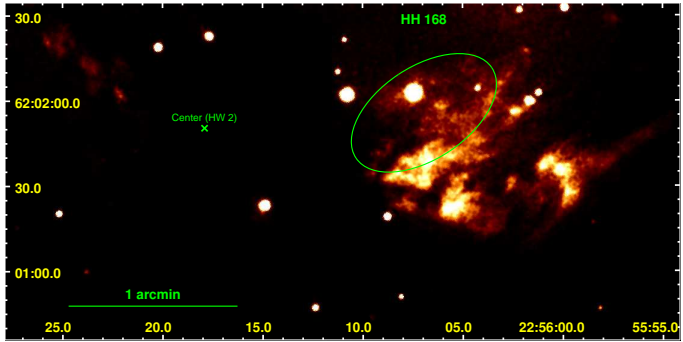}
   \caption{SDSS$^\thefootnote$ DR7 \citep{SDSS_DR7} i-band (7481 \AA) image of the Cep~A region. The ellipse indicates the soft X-ray emission. The diffuse object south of the ellipse is a reflection nebula probably excited by a source close to HW~2 or by HW~2 itself \citep{Hartigan_1986}.\label{fig:RegionOverview} }
\end{figure}

Cepheus~A is the second nearest high-mass star-forming region at a distance of $\sim$ 730~pc \citep[][see Fig.~\ref{fig:RegionOverview} for an overview]{Johnson_1957,Crawford_1970}.
Three Herbig-Haro objects (HH~168, \object{HH~169}, and \object{HH~174}) are located around the center of Cep~A, where several distinct radio sources have been found \citep[named HW 1-9;][ HW~8 and HW~9 were detected by subsequent observations]{Hughes_1984}.
Because of the strong absorption towards the center of Cep~A \mbox{($A_V > 75$~mag)} radio observations are the most suitable for imaging this region. In the IR wavelength range \citep[e.g. 12.5~$\mu$m,][]{Cunningham_2009}, none of the radio sources has been clearly detected and the region appears dark in Fig.~\ref{fig:RegionOverview}.

For the HH~169 and HH~174 outflows, HW~2 is currently the most promising candidate for the driving source, while it is unclear which source drives the HH~168 outflow, the possibilities being HW~2 and HW~3c, as well as an isolated region of star formation located at the eastern end of HH~168 \citep[e.g.,][]{Cunningham_2009}.

\stepcounter{footnote}
\footnotetext{\texttt{http://www.sdss.org/}}

The Cep~A region also contains multiple molecular outflows. The extent of the large-scale outflows is on the arcmin scale with typical velocities on the order of a few 10 km$\,$s$^{-1}$. The multicomponent appearance of the outflow geometry can be interpreted in terms of dense condensations redirecting the outflow \citep[][and references therein]{Codella_2003,Hiriart_2004}, but multiple outflows or the evolution of a single outflow are also possible. The position angle of the northeast-southwest molecular outflow is consistent with the elongation of HW~2 at radio wavelengths.
Observations of the gas dynamics, in particular of the H$_2$ gas \citep{Hiriart_2004}, show a clear velocity gradient within the western H$_2$ structure (blueshifted at the eastern end and at approximately zero velocity at the western end).

The prominent HH~object 168 is located west of the central sources of Cep~A.
Its kinematic lifetime is on the order of several 1000~years.
The H$\alpha$, [S{\sc ii}], and [O{\sc iii}] images observed by the Hubble Space Telescope \citep[HST,][]{Hartigan_2000} show that in the region of HH~168 no single type of shock
can be responsible for the observed emission. In some of the bright regions, the location of H$_2$ emission downstream of the H$\alpha$ emission is indicative of a shock-type that heats the material in front of the shock slowly without dissociating the H$_2$ molecules.

Radio emission has also been observed from a few regions within the HH~object \citep{Hughes_1990}. The radio brightest one is designated HW-object (without any number) by \citet[][]{Hartigan_1985} and is located at the eastern end of HH~168. Four (or even five) small emitting regions have been resolved within this object \citep[][]{Rodriguez_2005}, all of them being time-variable on timescales of years. Furthermore, they have large proper motions (arguing against a stellar origin) increasing from east to west from 120~km~s$^{-1}$ to 280~km~s$^{-1}$.
The motion of these objects is directed westward which is approximately consistent with the direction of  HH~168 itself.

\citet{Rodriguez_2005} proposed that the luminosity changes of the individual sources might be caused by holes in a stream flowing around condensations and exciting the radio emission. These authors estimated an outflow speed of about 900~km~s$^{-1}$, which still accelerates the radio-emitting objects.
The spectral indices of these radio emitting objects within the HW~object differ, pointing to different production mechanisms of the radio emission although simultaneous observations, as needed for time-variable sources, have not been carried out. One object seems to bare an ultra-compact, optically thick core, while another is consistent with shocked gas emission \citep{Garay_1996}.

In this region, \citet{Hartigan_2000} observed nine, approximately point-like, H$\alpha$ emission components, which they related to young T~Tauri stars, but \citet{Cunningham_2009} found no point-like sources in their infrared images in this region, even though their sensitivity should be sufficient.

Using H$\alpha$ images of HH~168, \citet{Lenzen_1988} measured the proper motions of individual knots ranging from 100 to 210~km~s$^{-1}$ (the HW-object in total has a velocity of 110~km~s$^{-1}$).  
\citet{Raines_2000} measured the HW~object in [Fe II] 1.644~$\mu$m and found an object moving with a space-velocity of $\sim$850~km~s$^{-1}$, while the bulk of the [Fe {\sc II}] emitting material moves only with 400~km~s$^{-1}$.
The H$\alpha$ lines of the HW~object, as well as the nearby knot~E, extend from zero to -400~km~s$^{-1}$ \citep{Hartigan_1986}.

\citet{Cunningham_2009} proposed that the bright $H_2$ emission within HH~168 is caused by high-velocity material encountering already decelerated material, thus forming shocks (shock speeds possibly reaching  400~km~s$^{-1}$). These authors further propose that the HW~object is produced by two colliding winds, one producing the HH~168 object that is driven by a source close to HW~3c and the outflow of HW~2 that drives the eastern HH~object.

In summary, there is no consensus about the nature of the observed objects in the literature. However, the large proper motion of some of those objects is sufficient for the production of X-rays by shocks.

The region of \object{Cep A} was observed with \emph{XMM-Newton} by \citet{Pravdo_2005} for 44~ks. They detected very soft diffuse X-ray emission at the position of HH~168 and constrained the total luminosity and temperature of the region. Due to the limited spatial resolution, they could not identify the precise location of the soft X-ray source(s).
The diffuse X-ray emission HH~168 shows a temperature of $T=5.8^{+3.5}_{-2.3}\times 10^6$~K  absorbed by a column density of $n_H = 4\pm4\times 10^{21}$~cm$^{-2}$.

\section{Observation and data processing}
The Cep A region was observed on 8 April 2008 for 80~ks with ACIS-I (ObsID: 8898). We used CIAO~4.1.2 to analyse the data, following closely the science threads published at the CIAO website \footnote{\texttt{http://cxc.harvard.edu/ciao/}}.
We used the archival data without reprocessing the $evt2$ file; therefore, the standard parameters are applied to the data. The analysis is restricted to events with energies between 0.3~keV and 10.0~keV, i.e., the energy range with a reliable calibration.

For both, detection and spectral analysis, an estimate of the background is needed.
The background for the X-ray emission related to HH~168 was estimated from two regions. The smaller one is a nearby ellipse of approximately the size of HH~168 ($a=30$\arcsec, $b=16$\arcsec) containing 557 events in the full energy range (30 photons in the 0.3~--~1.5~keV range, i.e., the range in which shock-induced X-rays are expected). The second region is more than seven times larger (60\arcsec ~ radius) and is located further off-axis to ensure that it is source-free, e.g., that no source is visible in the 2MASS images nor identified by any of the CIAO source-detection algorithms in that region); it contains 261 photons in the 0.3~keV~--~1.5~keV range. Both background estimates agree approximately in terms of their predicted count levels ($5.7\times10^{-3}$~cts/pixel for the large area, and $5.0\times10^{-3}$~cts/pixel for the smaller one, in the 0.3~--~1.5~keV energy range). We use the average of both values to determine the estimated background.
Their full range median energies coincide to within about 100~eV.

The spectral analysis was carried using XSPEC \citep{XSPEC} assuming an absorbed optically thin plasma emission model \citep[APEC][]{APEC}. We adopt $1\sigma$ errors throughout the paper.

\section{Results\label{sect:HH168_results}}
\begin{figure}
  \centering
   \includegraphics[width=0.49\textwidth]{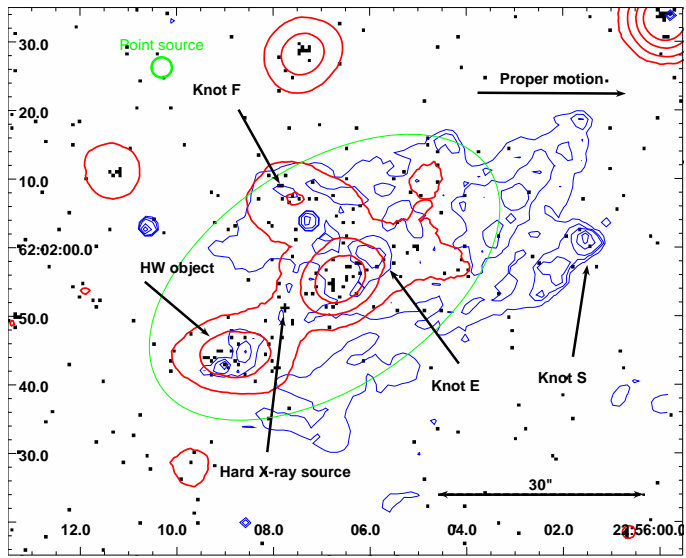}
   \caption{The region around HH~168 in soft X-rays (energies between 0.3~keV and 1.5~keV). Single pixels show individual photon positions. The blue contour is the H$\alpha$ emission from \citet{Hartigan_2000}, while the red contour shows the smoothed X-ray photon distribution which are multiples of four times the background rate. The proper motion of the HH object is directed toward the right.}\label{fig:HH168overview}
\end{figure}

Figure~\ref{fig:HH168overview} shows an overview of the HH~168 region in soft X-rays.
In soft X-rays, \texttt{vtpdetect} of the {\it Chandra} CIAO~4 tools finds an extended source in the HH~168 region. We indicate the excess region by the green ellipse in Fig.~\ref{fig:HH168overview}, which approximately coincides with the lowest X-ray contour (red). In this region, there is an excess of 101$\pm$13 photons over the expected background in the 0.3~keV -- 1.5~keV range while no diffuse excess is present in hard X-rays (1.5~keV~--~9.0~keV; see Table~\ref{tab:HH168_results} for the number of detected counts and estimated background counts).

\begin{table}
\caption{Diffuse X-ray emission in HH~168 (0.3~--~1.5~keV).}\label{tab:HH168_results}
\centering
\begin{tabular}{l c c c c}
\hline\hline
Component & Photons & Est. background & Median & Area\\
 &  & photons & energy & arcsec$^2$\\
\hline
Ellipse & 133 & 32.0 & 0.95 keV & 1463\\
HW object & 16 & 1.3 & 1.1 keV & 61\\
Knot E & 30 & 2.4 & 0.85 keV & 108\\
Knot F & 11 & 1.0 & 1.1~kev & 44\\
\hline
\end{tabular}
\end{table}

\begin{figure*}
  \centering
   \includegraphics[width=0.3\textwidth]{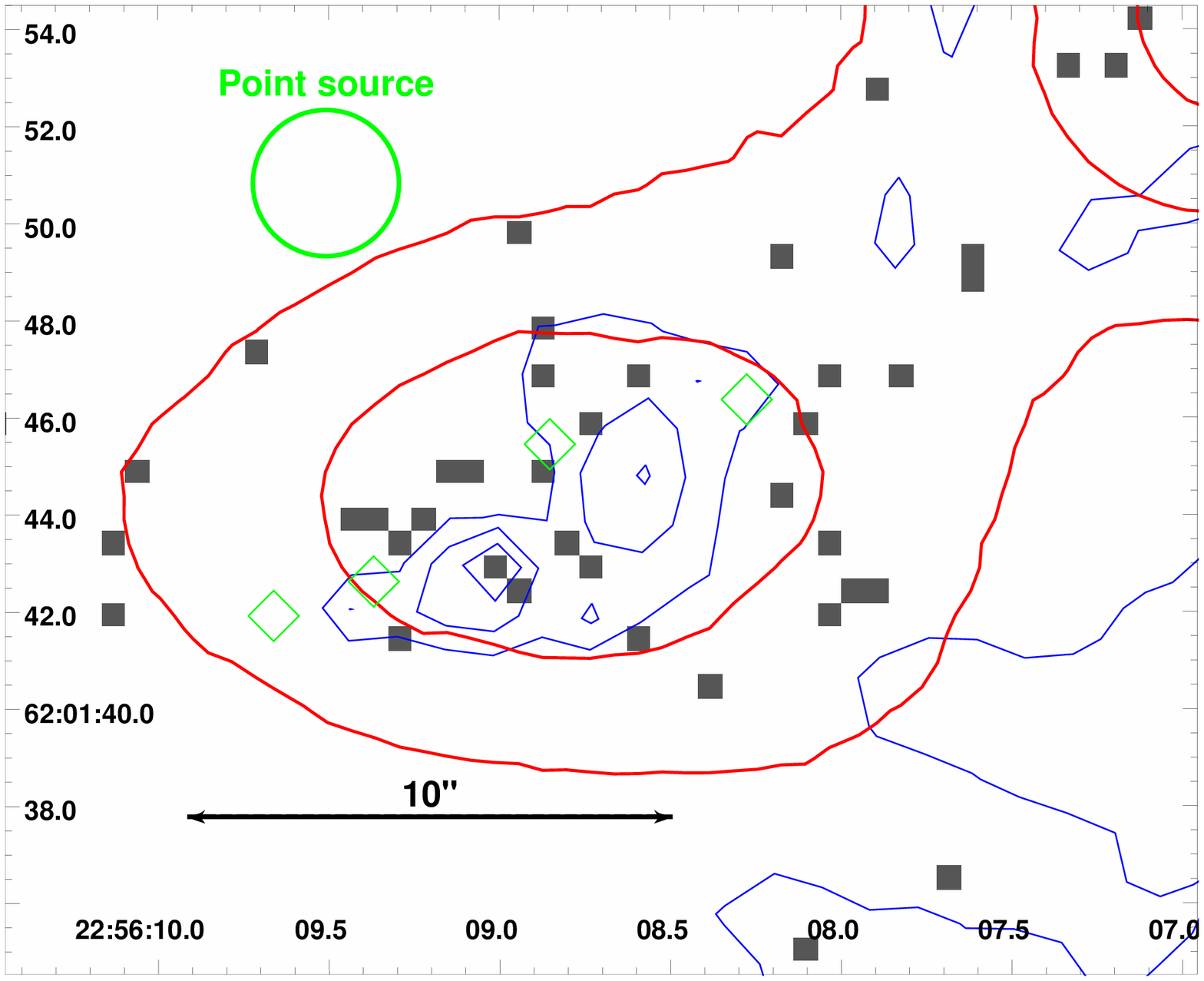}
   \includegraphics[width=0.3\textwidth]{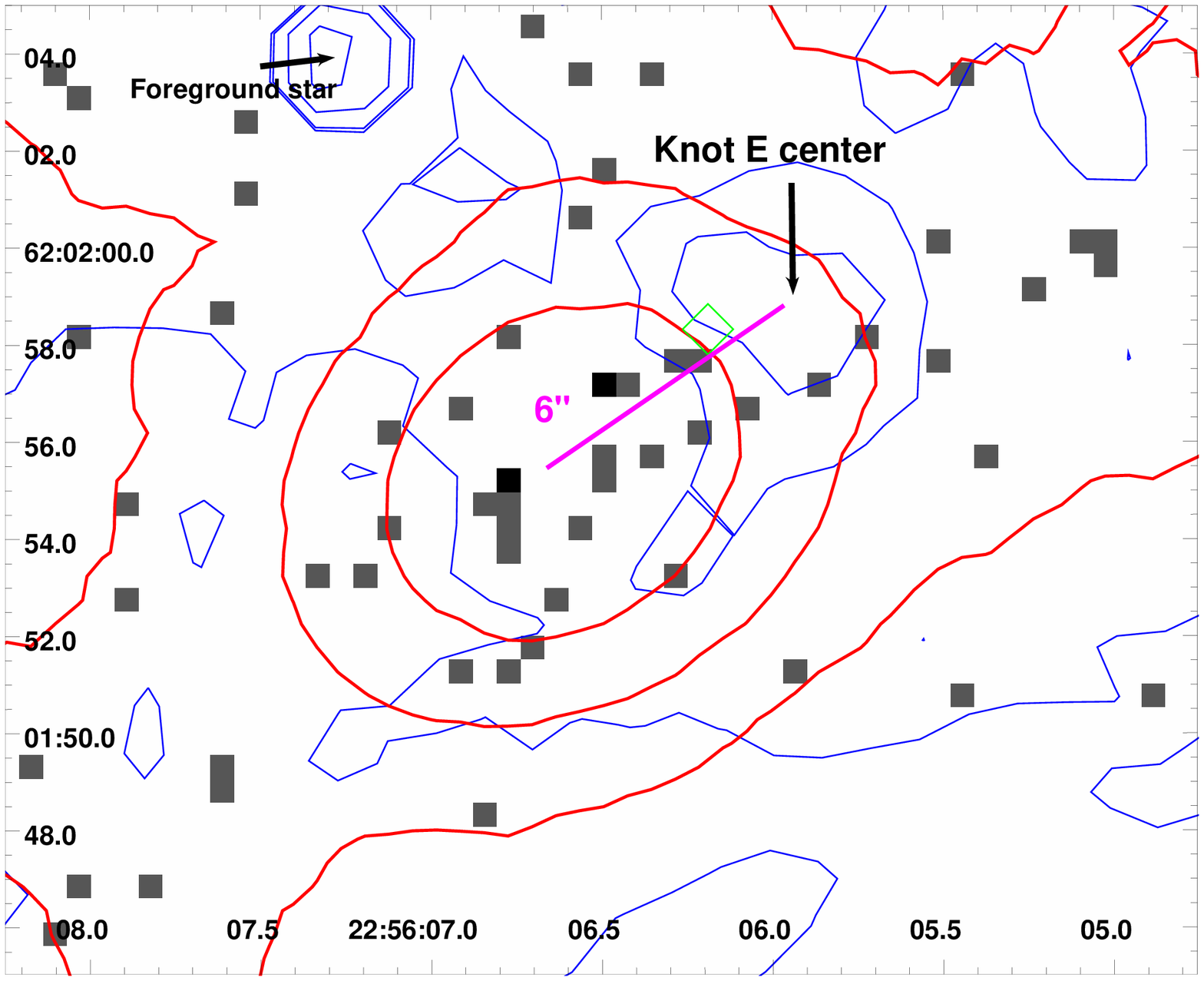}
   \includegraphics[width=0.3\textwidth]{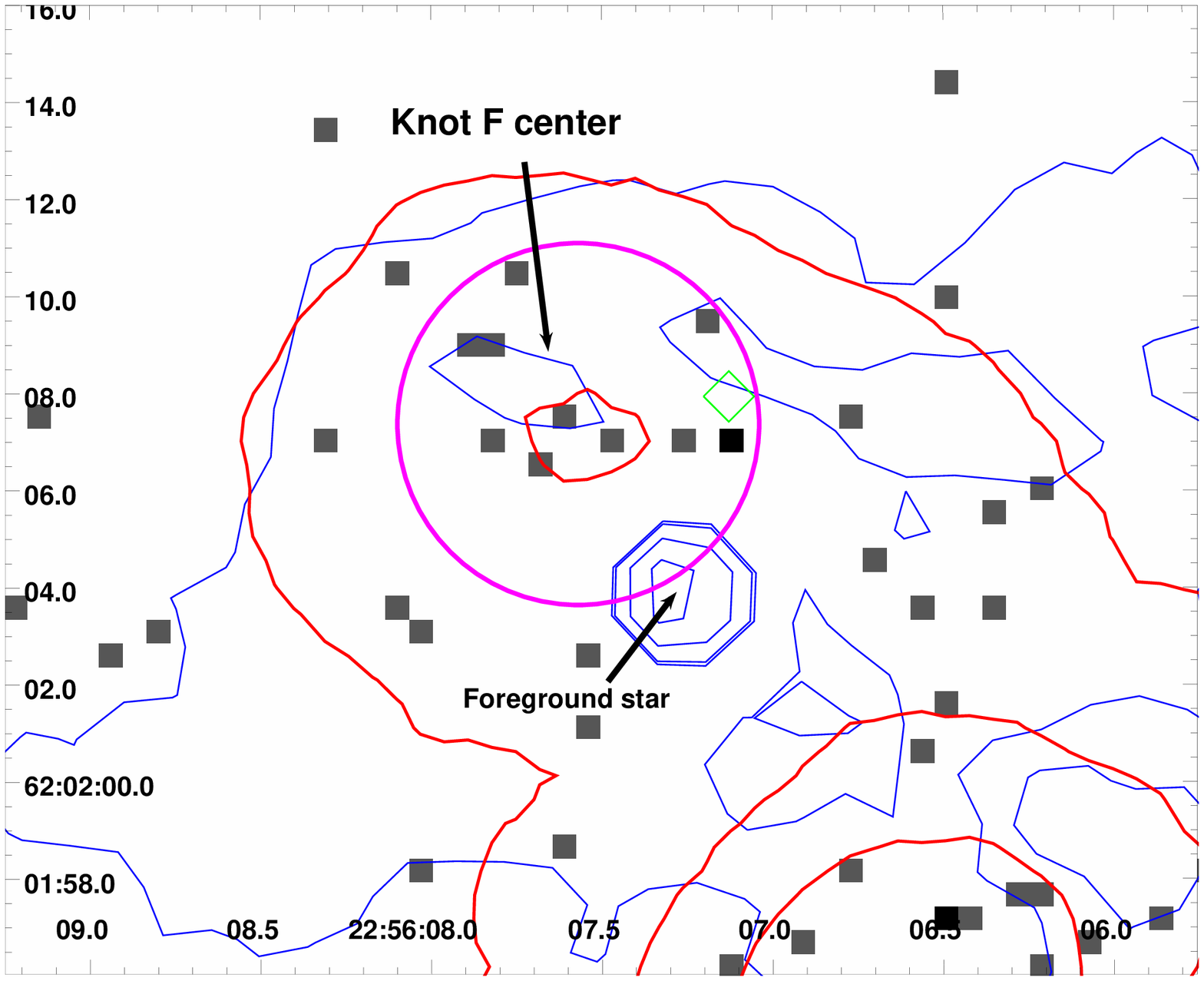}
   \caption{Zoom into selected regions. In the left panel, the HW-object, in the middle panel, the region around knot~E, and in the right panel the knot~F region are shown. Squares denote individual photons. The red contours are the X-ray contours from Fig.~\ref{fig:HH168overview}. H$\alpha$ emission is shown as blue contours and green diamonds designate radio sources from \citet{Rodriguez_2005,Hughes_1990}.}\label{fig:HH168closeup}
\end{figure*}

Within HH~168, approximately between the HW~object and knot~E, a hard, point-like X-ray source containing 10 photons is present (J(2000)=22:56:07.8 +62:01:51.3). It can be characterised by a median energy of 2.6~keV and no X-ray photons below 1.5~keV. The estimated unabsorbed X-ray luminosity is $7.5\times10^{29}$~erg/s assuming a plasma temperature of 1.4~keV and an absorbing column density of $3.5\times10^{22}$~cm$^{-2}$. This model reproduces the median energy.

\subsection{Morphology}
The soft X-ray contour (starting at four times the background level) overlaps with the H$\alpha$ contour, but lags behind the head of the H$\alpha$ emission (e.g., knot~S) by 20\arcsec. At the interface between the outflow and the ambient medium, only a very weak X-ray excess close to knot S is found. Within a circle of 5\arcsec~ radius around knot~S (the only H$\alpha$ knot showing bow-shock characteristics), we detect 5 photons, where 2 are expected from the background.

No soft point-like source can be detected at a level above 1.1$\sigma$. At this detector position, one would expect that more than 95\% of the photons from a point-source (with the observed soft spectrum) are located within a circle of  radius of 1.5\arcsec~ (in Figs.~\ref{fig:HH168overview} and \ref{fig:HH168closeup}, we show these circles for comparison). The area with the highest photon density in that region contains six photons within the 1.5\arcsec~ radius circle and lies within the highest contour in the middle panel of Fig.~\ref{fig:HH168closeup}.

Three individual components with an enhanced diffuse emission are clearly visible within the large region, and are also indicated in Fig.~\ref{fig:HH168overview} and shown in greater detail in Fig.~\ref{fig:HH168closeup} (Table~\ref{tab:HH168_results} summarizes their properties). The location of the eastern one coincides with the HW-object found by \citet{Hughes_1984} at cm-wavelengths, and the central one is located close to but offset to the east, the bright H$\alpha$ knot~E in the nomenclature of \citet{Hartigan_2000}, which shows also weak radio emission. The third component is located close to knot~F, which also emits at radio wavelengths.
There are further radio emission components at the western end of HH~168 showing equally strong radio emission but outside the X-ray contour of Fig.~\ref{fig:HH168overview}. On the other hand, all three regions with enhanced X-ray emission have radio components in their vicinity.

The low count density renders an estimate of the true size of the emission regions impossible.
For the diffuse emission close to the HW-object and knot~E, circular regions coinciding approximately with the second contour  in Fig.~\ref{fig:HH168closeup} ($4\times10^{-2}$~cts/pixel, eight times the background level) were used to extract the photons of these two components. For knot~F, we show the circle in Fig.~\ref{fig:HH168closeup}.

The knot~E region has an excess of 28$\pm$6 photons (area 108 arcsec$^2$) and the region around the HW object has an excess of 15$\pm$4 photons (area 61 arcsec$^2$). Close to knot~F, the X-ray emission has an excess of 10 photons (area 44~arcsec$^2$).
All of these regions are \textit{not} compatible with single point-sources as we would expect only 2-3 photons outside a 1.5\arcsec~ radius for the numbers at hand, in contrast to the observational result. However, it is possible that e.g., six of the photons in the knot~E region are indicative of an active pre-main-sequence star.
The emission close to knot~F does \textit{not} coincide with the position of the foreground star from which no X-ray emission  is observed.


There is diffuse emission beyond the three individual emission components. Excluding the knots from the region defined by the ellipse, a clear excess of 48$\pm$10 photons is found. We therefore regard the contours in Fig.~\ref{fig:HH168overview} as real, although we do not claim that the shape of the contours represent the precise geometry of the emitting volume.

\begin{figure}
  \centering
   \includegraphics[height=0.49\textwidth, angle=270]{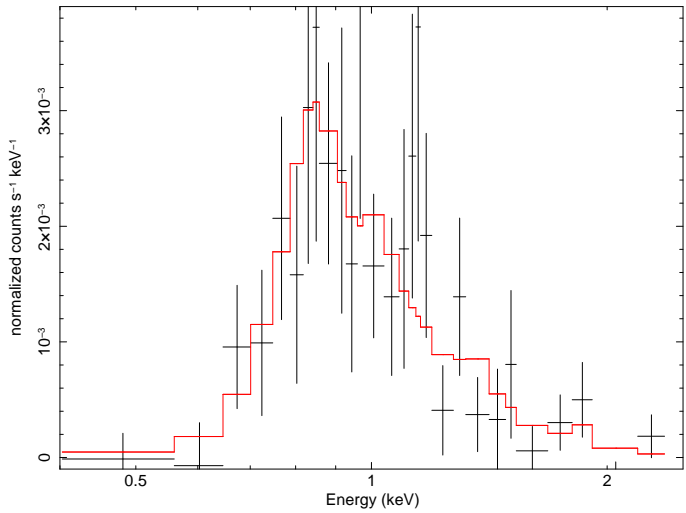}
   \caption{Spectrum of the photons within HH~168. The fitted model is shown in red. }\label{fig:HH168spec}%
\end{figure}

\begin{figure}
  \centering
   \includegraphics[width=0.49\textwidth, angle=0]{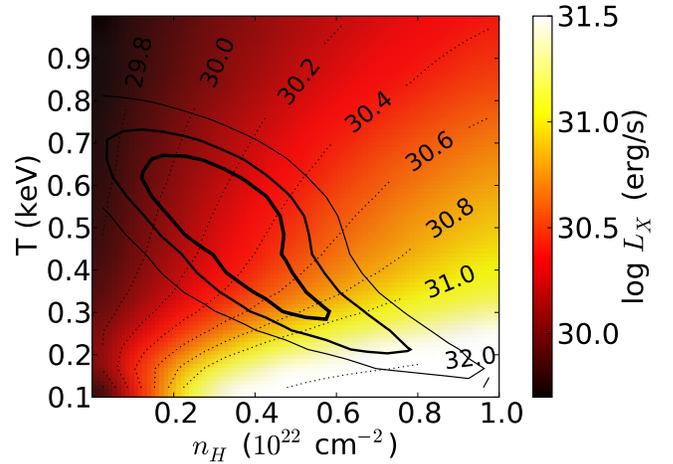}
   \caption{Contours showing different confidence levels of the plasma parameters. The contours designate the 1~$\sigma$, 90~\%, and 99~\% confidence levels. The X-ray luminosity is color coded and indicated by the dotted lines.}\label{fig:HH168banana}%
\end{figure}

\subsection{Spectral properties and energetics}
Considering the energies of the photons within these three regions, we find slight differences between the median energies of the photons within the HW-object (1.1~keV) and the knot~E photons (0.9~keV). A KS-test infers a probability of only 15\% that both samples are drawn from the same distribution. On the other hand it is probable (60\%) that the photons around knot~E and those of the large-scale diffuse emission are drawn from the same distribution.

Ignoring for a moment this difference, we show in Fig.~\ref{fig:HH168spec} the spectrum of the photons within the ellipse.
The spectrum can be described well by an absorbed, thermal, optically thin plasma with $kT=0.55\pm 0.1$~keV and an absorbing column density of $n_H = 3\pm1\times10^{21}$~cm$^{-2}$.
These values are consistent with the \emph{XMM-Newton} data of \citet{Pravdo_2005}.  Using the conversion between $n_H$ and $A_V$ from \citet{Vuong_2003}, we find that $A_V \sim 2$.

Using the best-fit values, the total unabsorbed energy-loss of the diffuse emission component in X-rays is about $1.7\times10^{30}$~erg~s$^{-1}$. Figure~\ref{fig:HH168banana} shows the contours of several significance levels in the ($kT$, $n_H$)-plane; this plot illustrates the dependence of the luminosity on the adapted spectral parameters.
The energy-loss within the smaller components scales with their photon numbers ($L_X \approx N/N_{total} \cdot L_{X}^{total} = 5\times10^{29}$, $3\times10^{29}$ and $2\times10^{29}$~erg$\,$s$^{-1}$ for knot~E, the HW~region, and knot~F, respectively).

The difference in the median energy of the photons can either be caused by a different absorbing column density or by an intrinsically different temperature of the plasmas. In either case, one needs to approximately double either $T$ or $n_H$ to increase the median energy by 0.2~keV.


An estimate of the electron density $n_e$ of the X-ray emitting plasma can be given by the volume emission measure ($EM$, provided by the fit) needed to achieve the observed flux by 
\begin{equation}
n_e = \sqrt{\frac{EM}{0.85 \cdot f\,V}}\, ,
\end{equation}
where $V$ is the volume of the emitting region, $f$ the filling factor, and the factor $0.85$ reflects the solar ratio of electrons to H-atoms in an ionized plasma).
The EM for the large region has a value of $6.4\times10^{52}$~cm$^{-3}$, while it is a few $1\times10^{52}$~cm$^{-3}$ for the individual regions.
Assuming spherical/ellipsoidal volumes, we find $4.2\times10^{52}$~cm$^3$ for the entire area, $1.1\times10^{51}$~cm$^3$ for the knot~E region, and $4.6\times10^{50}$~cm$^3$ for the HW~region leading to densities of $n_e = 1.3$~cm$^{-3}$ for the large-scale emission and about 10~cm$^{-3}$ for the smaller regions.

None of those objects exhibit significant variability in their light curves. However, the low number of counts is not sufficient to exclude minor variability (factor $\sim 2$) within 10~ks.

\section{Is the emission really diffuse?}
The appearance of a diffuse emission component can in principle be mimicked by a superposition of a large number of discrete sources as Cep~A is a young star-forming region.
We therefore assume in this section that a great number of YSOs produce the soft X-rays, and highlight the problems of this scenario. We concentrate on the 0.3~--~1.5~keV energy range to which the hard X-ray source does not contribute.
\begin{enumerate}
\item For the entire region of HH~168, not even one pixel contains three photons. Assuming that the individual sources contribute only 5 photons each, then 20 of these sources are needed. Lowering the significance of the source-detection algorithm as far as possible, only 26 sources are detected with a mean count number of about two.
Thus, a higher number of discrete sources ($> 50$) is required to explain the soft X-ray emission with point sources.
\item The relatively small absorbing column density should also ensure that pre-main sequence stars are detectable in the IR and possibly also as H$\alpha$ sources.
Taking the saturation limit of $L_X / L_{bol}\approx 10^{-3}$, we find that a single highly active mid-M dwarf would produce a single X-ray photon at a distance of 730~pc. Thus, about 100 M-dwarfs are needed within the HH~object to account for the total luminosity. 
A mid M-dwarf has an apparent V-magnitude of about 20 assuming that $A_V=2$. Thus, even in the unlikely case that only M-dwarfs and later objects are present in HH~168, at least some should have been detected with the available observations.
Furthermore, the X-ray emission does not seem to be concentrated in the shallow strip as the peaks in H$\alpha$. For the enhanced emission close to the knot~E, we note that the X-ray emission is offset from the bright H$\alpha$ emission and the potential position of the YSOs.
\item
No diffuse hard X-ray emission ($E > 1.5$~keV), expected to be present in low-mass pre-main-sequence stars, is observed. The median plasma temperature of the sources in the Orion nebula cloud \citep[COUP:][]{Preibisch_2005, Getman_2005} is more than a factor of three higher than the best-fit temperature of the diffuse HH~168 emission.
Taking the median count number of the COUP sources, the expected count number of this ``normal'' stellar source would be about 10~photons for the absorption and distance of Cep~A. The highest photon density within the HH~168 area contains six photons within the 95~\% region of a point source.
\item The probability of finding a YSO in the HH~168 region can be estimated using 
the stellar surface density of the Cep~A region derived by \citet{Gutermuth_2009} from mid-infrared imaging. Assuming a uniform distribution of their sources, we derive a probability of about 0.2~\% of finding one source within a 1 arcsec$^2$ region. The stellar surface density also matches approximately that of the Trapezium region in Orion, which was studied during the COUP project with a 25 times higher X-ray sensitivity than the observation of Cep~A. From the surface density of the 1616 COUP X-ray sources, we estimate a probability of about 0.5~\% of finding one X-ray source within 1 arcsec$^2$.
Both values correspond to less than one expected source within the three individual components, while two (five) sources are expected within the large source ellipse. These numbers are too low to account for the luminosity of the object.
\end{enumerate}

We regard a stellar population of only M-dwarfs and later objects without any hard X-ray emission, as an unlikely source of the observed diffuse X-ray emission. The shape of the emitting region (located within the H$\alpha$ contour) and the apparent softening of the outflow from east to west, are not explained. However, a limited number of weak discrete sources cannot be excluded.
We therefore conclude, that the vast majority of the emission is indeed diffuse.

\section{Are the X-rays produced in shocks?}
A solution for the heating of the observed hot plasma is shock heating. In this section, the properties of the observed X-ray emission are related to those expected from shocks.
\subsection{X-ray emission from shocks}
The observed X-ray emission is produced by plasma with temperatures of around or above $10^6$~K. 
Because shocks are the ultimate origin of this hot plasma (ignoring possible contributions by coronal emission of unresolved stars in this region) we recall some relations between properties observable at other wavelengths and X-rays.

From the strong shock jump condition, the temperature $ T$ of the plasma is
\begin{equation}\label{Eq:shockT}
T \approx 1.5\times10^5 \mbox{K} \left( \frac{v_{bs}}{\mbox{100~km~s}^{-1}} \right)^2\,,
\end{equation}
where $v_{bs}$ is the velocity of the pre-shock material in the shock rest-frame
and the cooling distance $d_{cool}$ can be found by the interpolation formula of \citet{Heathcote_1998} 
\begin{equation}
d_{cool} \approx 2.2 \times 10^{16} \mbox{cm}^{-2} n_0^{-1} \left(\frac{v_{bs}}{\mbox{100~km}\,\mbox{s}^{-1}} \right)^{4.5}\,,
\end{equation}
which depends on the pre-shock particle number density $n_0$ and has a $\sim$20\% accuracy within the 150-400~km$\,$s$^{-1}$ range \citep{Raga_2002}.
Furthermore, the mass flux rates for quasi-stationary shock configurations can be calculated from the thermal energy $kT$ and the $EM$ according to \citet{Guenther_2009} to be
\begin{eqnarray}
\dot M_{\rm shock} & \approx & 2.7\cdot 10^{-11}\frac{M_{\sun}}{\textnormal{yr}} 
              \left(\frac{EM}{10^{52}\textnormal{ cm}^{-3}}\right)
              \left(\frac{0.33\textnormal{ keV}}{\textnormal{k}T}\right)^{1.75} \; .
\label{eqn_mdot}
\end{eqnarray}

\subsection{Are the spectral properties consistent with shocks? \label{sect:coolingTime}}
The range of possible plasma temperatures (see Fig.~\ref{fig:HH168banana}) reaches from 0.1~keV to 0.6~keV. The corresponding shock speeds are 280~km~s$^{-1}$ and 680~km~s$^{-1}$, respectively; speeds in that range have been deduced in this region. The proper motion of individual knots is $\sim 300$~km~s$^{-1}$ and possibly even higher \citep{Raines_2000}, the emission line width might be as great as 400~km~s$^{-1}$ \citep{Hartigan_2000} and \citet{Rodriguez_2005} speculate about inhomogeneities in the outflow moving with 900~km~s$^{-1}$. A superposition of flows with different shock speeds might also result in the observed spectrum.

The observed low absorption is compatible with the detection of NIR and H$\alpha$~emission in that region. Differences in the absorbing column density have not been investigated with the available observations, although, a uniform absorbing column density is unlikely.

The derived densities (assuming a filling factor $f$ of unity) are about a factor 1000 lower than those estimated from optical emission lines \citep{Hartigan_2000}, but the thermal pressure $P$ of the hot plasma ($P=nkT$, $k$ Boltzmann's constant) and that of the optically observed material ($T\sim10^4$~K and $n \sim 10^4$~cm$^{-3}$) differ by only a factor of ten. To avoid the low pressure of the X-ray plasma, a filling factor $f$ of less than a tenth is mandatory.

Alternatively, the hot ions might represent a co-spatial component of a denser and cooler gas, which is not yet thermalised. In collision-dominated plasmas, a few mean free path lengths are sufficient to arrive at a Maxwellian energy distribution. The mean free path $l$ for collisions with neutrals of number density $n$ is
\begin{equation}
l=(n \pi a_0^2)^{-1}\approx 10^{16} n^{-1} \mathrm{ cm}^{-2}\,,
\end{equation}
where $a_0$ is the Bohr radius of the hydrogen atom \citep{Lang_1999}. For typical densities of $n=10^3$~cm$^{-3}$, a hot ion population would thermalise within a few AU. Thus, we cannot expect to find two distinct populations in a given volume in the abscence of heating and we reject this scenario.
Therefore, either the X-ray emission is produced in only a thin surface layer of gas filled bubbles, possibly at the interface between the driving flow and the ambient medium, or in smaller individual knots such as the H$\alpha$ or radio features within HH~168 \citep{Hartigan_2000, Rodriguez_2005}.

With an estimate of the density of the X-ray emitting plasma, we can derive the cooling time of the material by
\begin{equation}
\displaystyle \tau = \frac{3 kT}{n_e \Lambda(T) }  = 1.6\times10^{13} n_e^{-1} \mbox{s cm}^3 = 5 \times 10^5 \left( \frac{n_e}{\mbox{cm}^3} \right)^{-1}\mbox{yr}\,,
\end{equation}
where we assumed for the cooling function $\Lambda(T) = 8\times10^{-23}$~erg~cm$^3$~s$^{-1}$ and the best-fit model value $T=0.55$~keV.
Thus, only for densities of $n_e \sim 10^3$~cm$^{-3}$ are the cooling time and the kinematic lifetime of the outflow of the same order, since typical timescales for the age of the outflow are in the range of several 1000~years.

Only for densities above $n_e\sim 10^5$~cm$^{-3}$ is the cooling time sufficiently short for the plasma to remain within around 730~AU (an arcsec) of its heating location assuming a proper motion of 0.1\arcsec$\,$yr$^{-1}$. Otherwise, the timescale for the cooling is so long that the position of the plasma is not necessarily close to the location where the heating occurred.

If we assume that the HW~object and the knots E and F are currently heated by ongoing shocks, so that we observe the post-shock cooling zone, we can calculate the mass loss according to Eq.~\ref{eqn_mdot}, as $\approx 10^{-10} M_{\sun}$~yr$^{-1}$.

The total X-ray luminosity of HH~168 is an order of magnitude larger than the X-ray jet of DG~Tau (which is a pre-main-sequence star of less than one solar mass) and in-between those of HH~2 and HH~80/81. In those cases, the mechanical energy of the outflow is by far sufficient to power the X-ray emission, which is probably also true for HH~168; although the kinetic energy of the outflow is not known, that of the molecular outflow alone exceeds the required energy by a few orders of magnitude.

Thus, we conclude that all spectral properties are consistent with shock heating and the properties derived from other wavelengths.

\subsection{Comparison with HH~2 and HH~80/81}
Soft X-ray emission within similar HH~objects has also been detected from HH~2 \citep{Pravdo_2001N}, HH~80/81 \citep{Pravdo_2004}, and HH~210 \citep{Grosso_2006}. Their temperatures \mbox{($kT\sim0.1$~kV)} and association with an HH~object are very similar to HH~168. The density \mbox{$n_e \approx 50$~cm$^{-3}$} of the X-ray emitting material is for HH~2 and HH~80 higher than the values of HH~168. In these cases, the brightest X-ray emission components are correlated with bright H$\alpha$ emission knots exhibiting bow-shock-like structures.

Therefore, the heating most probably occurs in these objects because of an interaction of the outflow with another medium. This might be the ambient medium or a low-density bubble produced by a protostellar object as suspected by \citet{Pravdo_2004} because of the presence of a hard X-ray source between the soft X-ray emission components. 
For HH~168, the X-ray emission is not located at the head of the outflow where bow-shock structures are observed \citep{Hartigan_2000}, but a hard X-ray source within in the soft diffuse X-ray emission is also present.
The large-scale diffuse emission found in the case of HH~168 is not present in any of the other sources, which are extended on the scale of a few arcsec.

For HH~2 and HH~80/81, the X-ray emission correlates approximately with radio emission knots in the sense that a radio component is located close to the X-ray emission, while there are also radio knots without X-ray emission. Knot~E in HH~168 might be an exception of this rule.

\section{Location of shocks}
Where should we expect the shocks heating the plasma to the observed X-ray emitting temperatures to be located?
Because of the long cooling time of the X-ray emitting material, the observed X-ray emission might have been heated some time ago.

\subsection{Shocks at the head of the outflow}
The natural position of strong shocks is the head of the outflow since there the velocity difference is largest, but the bulk of the X-ray emission is located in the eastern part of the H$\alpha$ emission (see the X-ray contours in Fig.~\ref{fig:HH168overview}). If HH~168 is bounded by a shock front sufficiently strong to reach X-ray emitting temperatures, only a small offset is expected, because some time is needed to reach the ionisation equilibrium. For a density of $10^3$~cm$^{-3}$, we checked the timescale with a shock code that explicitly calculates the ionisation timescale \citep{Guenther_2007} and find, that the shock would produce \ion{O}{vii}, whose line emission dominates plasmas of the observed temperature, within 30~AU. The calculation scales inversely with the electron density so that even for a lower electron density of a few 10~cm$^{-3}$, the ionisation equilibrium is reached after less than thousand AU (less than one arcsec). This value is much smaller than the observed offset between the X-ray emitting material and the H$\alpha$ contour.

The brightest H$\alpha$ knot \citep[knot S, see][]{Hartigan_2000} shows the characteristic properties of a bow shock with the ambient medium, but has, at most, only weak excess X-ray emission (three photons above the estimated background). A stronger absorption at this part of the outflow than at the eastern end of HH~168 might reduce the observed flux.
Doubling the absorption would decrease the count rate by a factor of four rendering the detection of soft X-rays virtually impossible, but
a lower shock velocity or a smaller radius of the obstacle would also result in a smaller amount of material being heated to X-ray emitting temperatures and thus a reduction in the observed flux.

The shocks might have been more energetic in the past, so that in case of a low plasma density ($n_e \lesssim 10^3$~cm$^{-3}$) the observed material was heated much earlier in the outflow history and the luminosity at the current shock front (in particular knot~S) is below the detection limit.
The shape of the X-ray emission might then represent the ``shock history'' of the outflow. 

\subsection{Shocks within the HH-object}
Since most of the X-ray emission is observed within the outflow, another explanation for the observed features is presented.

There has been some speculation about a high velocity component ($\sim900$~km$\,$s$^{-1}$) within HH~168 \citep{Raines_2000, Rodriguez_2005}, which is sufficiently fast to produce X-ray emission within HH~168 when encountering denser and slower material, such as a thin bubble expanding from the hard X-ray source between the HW~object and knot~E similar to the scenario proposed by \citet{Pravdo_2004} for HH~80.
The individual X-ray components are not aligned in the east-west direction of the large-scale outflow, but a single precessing jet \citep[as proposed for the eastern HH~objects,][]{Cunningham_2009} with an opening angle of only 20 degrees can reach all X-ray and radio components (assuming a driving source at Cep~A East). A deflection of the outflow at high density clumps or different outflows might also explain the morphology.
Therefore, the same high velocity flow heating the HW-object might also heat the material close to knot~E and knot~F. 

The highest X-ray luminosity is observed close to knot~E
and might be a remnant of strong shocks in that region. The relatively low density of the X-ray emitting material and its lower median energy compared to the HW object, support a scenario in which the heating happened earlier and the currently emitting material is not at the position of its heating.

The strongest radio emission is detected at the HW-object, the region closest to the driving source of the HH~168 outflow, \citep[or harbouring the driving source itself, ][]{Garay_1996}. \citet{Goetz_1998} and \citet{Cunningham_2009} speculate that the outflow might be deflected somewhere around the HW-object, or that two outflows (those of HW~2 and HW~3c) collide at this position. It is possible that the heating of the plasma happens somewhere near or within the HW-object and the X-ray emission westwards is from cooling plasma.
In this scenario, the difference between the median energy of the HW-object and the knot~E photons reflects the cooling of the material, and a density of a few 100~cm$^{-3}$ is necessary to explain the softening of the photons by radiative losses.
The lack of significant X-ray emission at the western end of the HH-object would be naturally explained in this scenario. The weak X-ray emission close to knot~S is then, if real, caused by small shocks of the outflowing material with the ambient medium.

However, the correlation of the optically observed material (in particular H$\alpha$ emission), the radio emission and the brightest X-ray emission points to a similar heating mechanism for all of these features. The electron density estimated from the radio observations is about a hundred times higher than that estimated from X-rays but another factor of ten lower than calculated for the material radiating in forbidden emission lines. This would imply a pressure balance, if less than a tenth of the observed volume is filled with a hot plasma. Thus, the observed X-ray emission probably traces the (former) location of strong shocks that in the mean time decelerated explaining the offset between H$\alpha$ and X-ray emission seen in knot~E, and on a smaller scale within the HW~object. The diffuse X-ray emission, which is also present between the individual knots, is probably very thin plasma heated earlier in the outflow history and  cooling slowly.


%
%

\subsection{A reverse shock}
Another possible way of explaining the offset between the H$\alpha$ emission and the X-ray emission are reverse shocks. Some time ago, two shock fronts originated in the head of the outflow, one moving in the primary direction of the flow, and the other traveling backwards relative to the bulk material. At any given point in time, the reverse shock appears similar to internal working surfaces, which are caused by intrinsic inhomogeneities in the jet, but its post-shock cooling zone can be found towards the head of the outflow. In this case, the X-rays of the HW~object or knot~E are the sign of an extended cooling zone. At some distance to the shock, the matter cools to H$\alpha$ emitting temperatures. This scenario is attractive, because it can explain the observed offset between X-ray emission and H$\alpha$ contours, although, we would expect at least some optical emission close to the actual shock front.

\subsection{An expanding bubble filled with hot plasma}
For a cooling time on the order of the kinematic lifetime of the outflow and a launching point somewhere in the region of the radio sources in Cep~A East, it is possible that the X-ray emitting material might be heated closer towards the centre where the absorption is higher and thus the heating source may remain undetected in X-rays by current observations.

On larger scales ($l \gtrsim 1$~parsec), bubbles filled with a very thin ($n_e< 1$~cm$^{-3}$) and hot ($kT \sim 1$~keV) X-ray emitting gas have been found \citep[][]{Townsley_2003, Guedel_2008S}. 
The minimum electron densities in the case of HH~168 are close to these values. However, the proposed B~stars in Cep~A East have a lower mass-loss and slower wind speeds than the O~type stars usually needed to fill the large bubbles.
Furthermore, in all other cases the X-ray emitting gas seems to be channeled by surrounding H$_2$, while in this case, the $H_2$ emission shows signatures of strong interaction with the ambient medium \citep[bow shocks, ][]{Cunningham_2009} co-spatial with the X-ray emission. 
However, HH~168 might be a kind of a transition object, where a low density plasma, heated either within the outflow or close to the driving sources, survives in a bubble blown out by earlier outflow episodes.

\section{Summary}
The \textit{Chandra} observation of HH~168 confirms the presence of diffuse soft X-ray emission within HH~168 as seen by XMM-\textit{Newton}.
The spectral properties of both observations agree well.
The higher angular resolution of the \textit{Chandra} observation clearly shows the diffuse character of the emission and discards YSOs as the origin of the X-rays. Furthermore, the substructure within the X-ray emission shows a good correlation with the optical and radio ``knots''.
Although this correlation has been seen in other HH~objects such as HH~2 and HH~80/81, the bulk of the X-ray emission in the case of HH~168 is not at the location of bow shocks. Instead, the X-rays are displaced towards the driving sources against the outflow direction by several thousands of AU. We regard this as evidence of internal shocks probably powered by a high velocity component within the large-scale outflow. The X-ray emitting plasma correlates with H$\alpha$ emission but is displaced from the H$\alpha$ knots. This points, in combination with the low density of the X-ray emitting plasma, to cooling plasma heated by the same outflow component leading to the currently observed H$\alpha$ emission.
The detection of diffuse emission apparently unrelated to the knots is indicative of a thin hot plasma heated by earlier shock events.

\it{Note added in proof: After acceptance of this paper an article dealing with the same X-ray data appeared on astro-ph \citep{Pravdo_2009}. However, these authors concentrate on the central region of Cep~A. Their results concerning the spectral properties and the location of the X-ray emission of HH~168 are compatible with ours.}


\begin{acknowledgements}
      We thank the referee, Dr. M. G\"udel, for his valuable comments which improved the paper.
      This work has made use of data obtained from the $Chandra$ data archive and the SDSS (http://www.sdss.org/).
      P.C.S. acknowledges support from the DLR under grant 50OR0703.
      H.M.G. acknowledges support from the DLR under grant 50OR0105.
\end{acknowledgements}
\vspace*{-2mm}
\bibliographystyle{aa}
\vspace*{-4.5mm}
\bibliography{HH168}
\end{document}